# Relationships between different Macroeconomic Variables using Vector-Error Correction Model

*Saannidhya Rawat*

## I. INTRODUCTION

Societies in today's world experience huge volume of transactions, all taking place at an extremely high speed. This makes tracking the direction of the overall economy almost inestimable. Still, to understand the economy as a whole, economists have been diligently working to invent methods that could allow us to understand the characteristics of an economy and gauge the economic position of a country. Although these methods are far from perfect, they are nevertheless helpful in developing an understanding of the functionality of an economy and help the economists in anticipating in advance the events that have previously occurred when the economy behaves in a certain manner. To facilitate these methods focused on obtaining realistic models of the economy, economists use variables called economic indicators. Economic indicators are tools developed by economists to measure economic performance and also predict future performance of an economy. Although other sound metrics have been suggested by various studies, such as Index of Sustainable Economic Welfare (ISEW) by Stockhammer & Hochreiter (1997) or Genuine Progress Indicator (GPI) by Cobb, Halstead & Rowe (1995), Gross Domestic Product (GDP) remains the most important metric used when trying to gauge the performance of an economy. Broadly, GDP is defined as the total dollar value of all the finished goods and services produced within a country's territory during a pre-defined time period. Using GDP as a measure of a nation's economic health makes sense because it is essentially a measure of how much buying power a nation has over a given time period. GDP is also used as an indicator of a nation's overall standard of living because, generally, a nation's standard of living increases as GDP increases. A tool used by the Federal Reserve Bank (FRB) to manipulate the direction of the economy is the discount rate, which is the rate at which banks can borrow from FRB. When discount rate offered by Federal Reserve is low, the banks are able to borrow at a low cost from FRB, allowing them to give out loans in the market a cheaper rate, engendering the borrowers to take loans and boost economic activity in the market. The opposite effect is observed when discount rate is high. BE Hansen and Ananth Seshadri (2013) observe countercyclical relationship between productivity and discount rate in the short-run and surmised that these benefits are offset in the long-run. Consumer Price Index (CPI) is another economic indicator widely used today as a measure of inflation rate in the economy. CPI is obtained using a basket of goods. The price of this basket is determined and is then compared with the price of the basket of goods for a pre-determined base year. Mundell (1963) focused on inflation and how it can drive the real interest rate in the market. This seminal work in theory of inflation paved way for developing models that revealed relationships between cost of borrowing in the market and inflation rate, which is also related to discount rate offered by FRB (RH Gordon 1982). Federal Reserve Bank uses discount rate as a tool to fight the unwanted growth of inflation and to stymie the cost of borrowing in the markets.

All of this suggests that there must exist a relationship between different macroeconomic variables that represent the economy. Through this paper, an attempt has been made to quantify the underlying relationships between the leading macroeconomic indicators. More clearly, an effort has been made in this paper to assess the cointegrating relationships and examine the error correction behavior revealed by macroeconomic variables using econometric techniques that were initially developed by Engle and Granger (1987), and further explored by various succeeding papers, with the latest being Tu and Yi (2017). The question that is first asked and then answered is that what is the extent of inter-dependence within the economy. If we can arrive at an accurate estimation of this relationship, we can devise relevant policies that take into account the extended ramifications a policy might have, which would help in formulating more informed policies and facilitate long-term growth. To analyze this, vector-error correction model is designed using four macroeconomic variables, with three co-integrating relationships and 2 lagged vectors for the examination of relevant relationships. Gross Domestic Product, Discount Rate, Consumer Price Index and population of U.S are representatives of the economy that have been used in this study to analyze the relationships between these economic indicators and understand how an adverse change in one of these variables might have ramifications on the others. This is performed to corroborate the belief that a policy maker with specified intentions cannot ignore the spillover effects caused by implementation of a certain policy.

## II.    DATA

For this study, most of the data were collected from Federal Reserve Economic Data (FRED), which is a databased maintained by the research division of Federal Reserve Bank of St. Louis. The data are in time series format, compiled by the Federal Reserve and collected by government agencies such as U.S Census Bureau and Bureau of Labour Statistics (BLS). Data were collected on a per quarter basis, with data starting from $1^{st}$ quarter of 1950 to the $1^{st}$ quarter of 2017, providing a total of 269 observations.

Table 1 - Source and Naming Conventions

| Description | Variable | Source |
| --- | --- | --- |
| Real Gross Domestic Product | gdp | Federal Reserve Economic Data |
| Discount Rate | disc_rate | Federal Reserve Economic Data |
| Consumer Price Index | CPI | Federal Reserve Economic Data |
| Population | us_pop | Federal Reserve Economic Data |

Gross Domestic Product has been taken in real terms, and is measured in terms of chained-dollars, which is a method of adjusting real dollar amounts for inflation over time, so as to allow just comparison of figures from different years.
Discount Rate, which in U.S is the interest rate charged to commercial banks and other depository institutions when they borrow from Federal Reserve, has been empirically known to have tremendous impact on the economy.
Consumer Price Index is formed by dividing the costs of a pre-specified market basket in a given year by cost of the same market basket in a base year. This permits the assessment of increase in prices compared to the base year of the same goods and services, for which we use the term

"inflation". CPI is a widely used statistic for gauging inflation or deflation and has been proven quite powerful in doing so.

Population of United States today is thrice of what it was a century ago. As the number of mouths that are needed to be fed increase, a surge in population also provides more people who can participate in various economic activities, thus, supporting the overall economic output produced. Hence, taking the number of people living in a country into consideration is deemed important because the size of a nation greatly impacts the direction in which the economy is heading.

The data for these variables were used to perform some explanatory data analysis so that better understanding of these metrics could be made possible. The table below provides the summary statistics for the data used for this study –

Table 2 - Summary Statistics for level and differenced variables

|  | Mean | Standard Deviation | Minimum | Maximum |
|---|---|---|---|---|
| Δ gdp | 55 | 66 | -315 | 235 |
| Δ disc_rate | 0.00 | 0.46 | -2.10 | 1.71 |
| Δ cpi | 0.8 | 0.83 | -5.0 | 3.3 |
| Δ population | 650 | 124 | 417 | 1426 |
| gdp | 8168 | 4533 | 2085 | 16903 |
| disc_rate | 4.5 | 2.8 | 0.5 | 14.0 |
| cpi | 108 | 74 | 24 | 244 |
| population | 237735 | 50190 | 150852 | 325108 |
| n.o of observations |  |  |  | 269 |

Table 3 - Results of Unit-Root Testing

| Variable | Test-Statistic | Order of Integration where Critical value @ 5% |
|---|---|---|
| gdp | 2.03 | I(1) |
| disc_rate | -2.27 | I(1) |
| cpi | 3.1 | I(1) |
| us_pop | -0.27 | I(1) |

*Graphical Analysis*

Graph 1- 8 Long-Run & Short-Run Variables

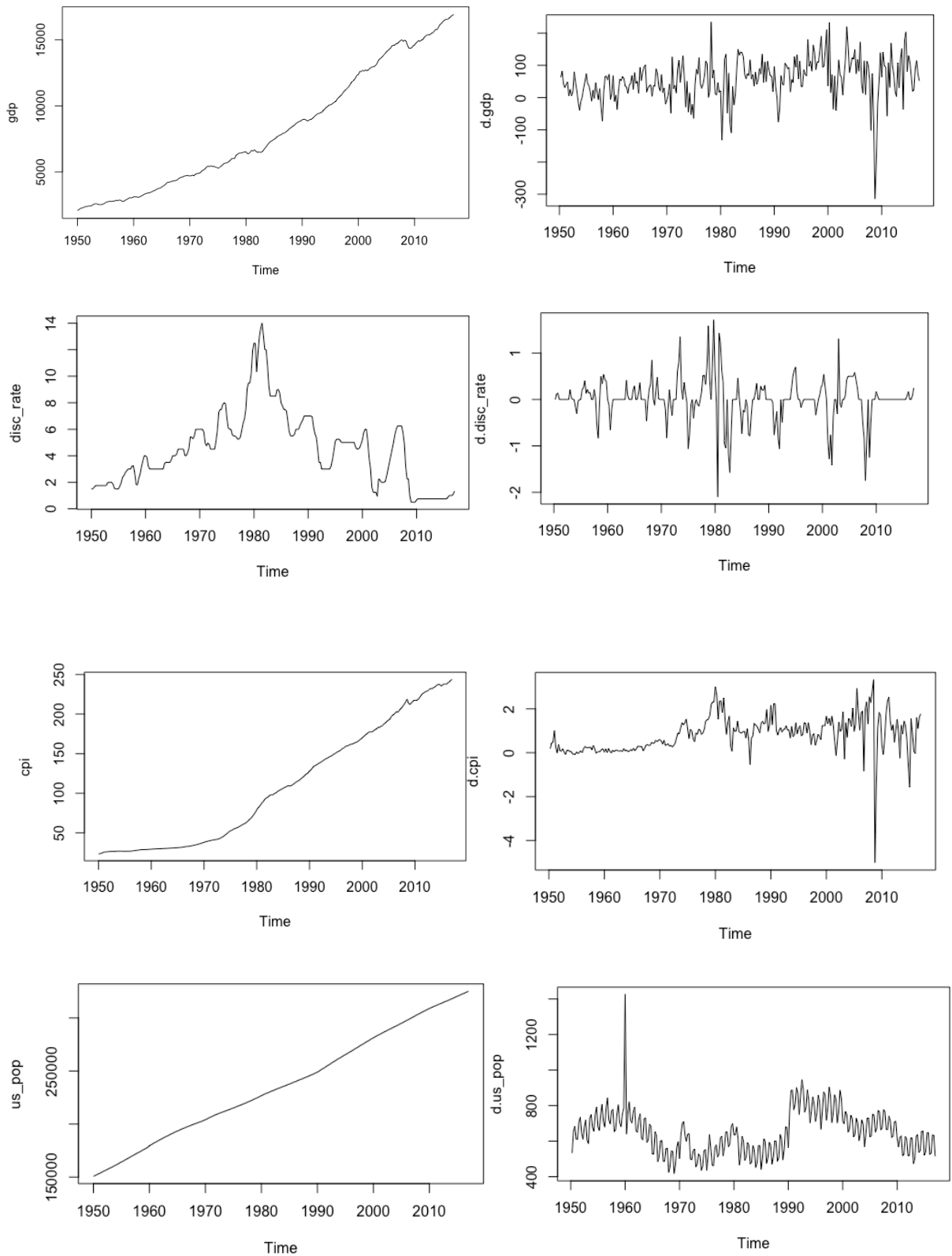

The graphs above confirm that the long run variables are non-stationary as they do not reveal mean-reversion. Thus, first difference of these variables was taken. After taking the difference, if the variables reveal mean reversion and pass the unit root test, then the level variables can be said to be I(1) and the differenced variables can be deemed I(0). This can be confirmed by observing the graphs presented above. Also, Table 3 statistically corroborates the claim that the above variables are I(1).

*Engle-Granger Cointegration Test*
For identification of cointegrating relationships, first Engle-Granger two-step method (1978) was following to check whether at least one relationship can be identified. Afterwards, the Johansen cointegration test was pursued to identify co-integrating systems based on vector autoregression (VAR) so that more than one cointegrating relationships can be identified. The results of Engle-Granger two-step method have been provided below to prove that there exists at least one cointegrating relationship. If this technique provides stationary residuals, in that case, the variables can be considered adequate for Error Correction Modeling and a VECM can be developed using these macroeconomic indicators.

Table 3 Engle-Granger two-step Method OLS results

====================================================
| | Dependent variable: gdp |
|---|---|
| disc_rate | -85.000*** |
| | (10.000) |
| cpi | 27.000*** |
| | (1.800) |
| us_pop | 0.050*** |
| | (0.003) |
| Constant | -6,232.000*** |
| | (415.000) |
| Observations | 269 |
| $R^2$ | 0.990 |
| Residual Std. Error | 393.000 (df = 265) |
| F Statistic | 11,806.000*** (df = 3; 265) |
====================================================

Graph 9 Residuals of the Model

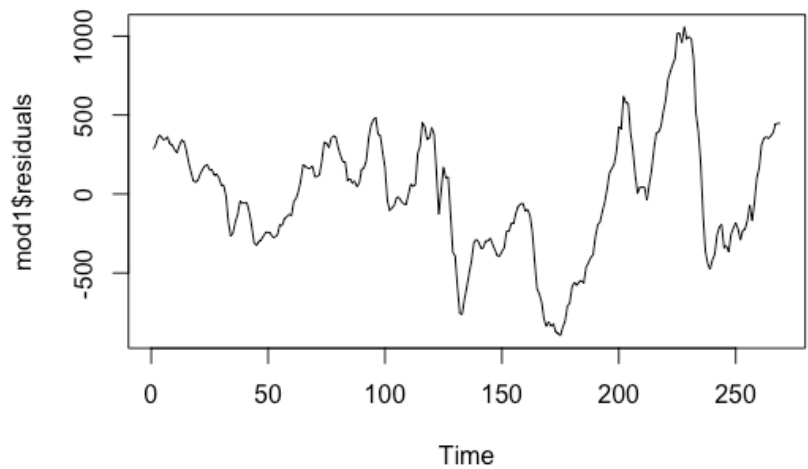

Table 4 - Unit-Root Testing Result

|  | **Test-Statistic** | **Order of Integration when Critical Value @ 5%** |
| --- | --- | --- |
| Residuals | -2.93 | I(0) |

The above summary output proves that a linear combination exists for the selected non-stationary variables such that when this combination is applied to these set of variables, the series taken together are stationary and hence, said to be co-integrated. This relationship can be used in our VECM model to define the error correction part and to gauge how quickly the variables return back to the steady-state or equilibrium.

In the next section, the functional form for the vector error correction model will be defined along with execution of Johansen cointegration test.

## III. EMPIRICAL METHODOLOGY

In this section, the functional form selected for this study will be explained. As mentioned before, Johansen's procedure for Vector Error Correction Models was used for this study. The key advantages of using this approach is that it can take numerous co-integrating relationships into consideration, all variables are treated as endogenous and testing for long-run parameters can be executed. The model that was considered for this study, in its generalized form, looks like the following –

$$\Delta y_t = \Pi \, y_{t-1} + \sum_{i=1}^{k} \beta_i \Delta y_{t-i} + \sum_{i=1}^{h-1} \alpha_i D_i + \varepsilon_t$$

where,

$\Delta y_t$ is a p x 1 vector, p with differenced variables

$\Pi$ is a p x p long-run coefficient matrix with co-integration relationships

$y_{t-1}$ is a p x 1 vector with variables in level form

$\beta_i$ are p x p short-run coefficient matrices for i = 1, 2, ….., k

$\Delta y_{t-i}$ is a p x 1 vector with differenced lags for i = 1, 2, ….., k

$\varepsilon_t$ is a p x 1 vector constituting stochastic terms where $\varepsilon_t \sim N(0, \sigma^2)$

$\alpha_i$ are p x p matrices for i = 1, 2, ……., h-1 coefficients for dummy variables

$D_i$ are p x 1 vector for i = 1, 2, ……., h-1 dummy variables

To determine the number of long-run relationships, Johansen's testing sequence (1990) was used, which says that if r is the rank of $\Pi$, 1 < r < g, where g represents full rank. For this model, r was identified to be 3.

Table 5 - Johansen's Testing Results

| Variable | Test-Statistic | Critical value @ 5% | Result |
| --- | --- | --- | --- |
| r <= 3 | 5.3 | 9.2 | Fail to reject the null |
| r <= 2 | 19.7 | 15.7 | Reject the null |
| r <= 1 | 40.5 | 22.0 | Reject the null |
| r = 0 | 64.2 | 28.1 | Reject the null |

To select the appropriate lag length for short-run variables, multivariate version of BIC was used, which can be defined as:

$$MSBIC = log|\hat{\Sigma}| + \frac{k'}{T}\log(T)$$

with k being the number of regressors

The adequate lag length determined via this approach was 2. Moreover, macroeconomic variables tend to reveal seasonal patterns. For example, GDP almost always increases in the 4$^{th}$ quarter at a higher rate than any of the other quarters. Hence, dummy variables were also introduced to capture seasonality.

In the next section, results using this approach are presented. Regression output, model performance results, Impulse Response Functions and Variance Decomposition Matrices will facilitate interpretation of the results and provide a better understanding of the underlying dynamics of the Vector Error Correction Model.

## IV. RESULTS

As the number of macroeconomic variables selected for this study were four, ultimately, four regression equations were estimated from the vector error correction model with each first differenced macroeconomic variable acting as the dependent variable in the system of equations. Johansen's method is powerful in the sense that it treats all the variables symmetrically, allowing all of them to be dependent variables simultaneously. This permits revelation of inter-dependent relationships that might not be easily extrapolated by classical linear regression techniques.
In the output above, all macroeconomic variables are treated as dependent variables.
Const represents the intercept and exo1, exo2 and exo3 are long-run co-integration relationships revealed in Table 7.
Results for each of these different dependent variables have been posted below –

Table 8 - d.gdp as the dependent variable

|  | **Estimate** | **Std. Error** | **T-Statistic** | **Pr(>|t|)** |
|---|---|---|---|---|
| d.gdp.l1 | 0.34225 | 0.06312 | 5.42 | <0.0001 *** |
| d.disc_rate.l1 | -15.24304 | 9.82121 | -1.55 | 0.12191 |
| d.cpi.l1 | -1.15313 | 5.70985 | -0.2 | 8.40E-01 |
| d.us_pop.l1 | -0.00473 | 0.05214 | -0.09 | 0.92779 |
| d.gdp.l2 | 0.24189 | 0.06458 | 3.75 | 0.00022 *** |
| d.disc_rate.l2 | -11.174 | 9.50808 | -1.18 | 0.24102 |
| d.cpi.l2 | -13.40367 | 5.76476 | -2.33 | 0.0209 * |
| d.us_pop.l2 | 0.01147 | 0.05227 | 0.22 | 0.82653 |
| const | 29.90084 | 23.05252 | 1.3 | 1.96E-01 |
| sd1 | 16.306 | 11.46183 | 1.42 | 0.15608 |
| sd2 | 4.54524 | 10.78304 | 0.42 | 0.67374 |
| sd3 | -0.62482 | 11.36326 | -0.05 | 0.95619 |
| exo1 | 0.00287 | 0.00172 | 1.66 | 0.09721 |
| exo2 | 0.00236 | 0.00176 | 1.34 | 0.18044 . |
| exo3 | 0.00176 | 0.00178 | 0.99 | 0.32452 |
| $R^2$ | | | | 0.257 |
| F-statistic | | | | 4.93 on 18 and 246 DF |
| Sample Size | | | | 265 |

In Table 8, with short-run GDP as the dependent variable, the first and second lags of short-run GDP are statistically significant at 0.1% significance level, along with $2^{nd}$ lag of CPI revealing correlation with GDP. Furthermore, exo1 is statistically significant at 10% significance level. $R^2$ is 0.257.

Table 9 - d.disc_rate as the dependent variable

|  | Estimate | Std. Error | T-statistic | Pr(>|t|) |
|---|---|---|---|---|
| d.gdp.l1 | 0.0015003 | 0.0004254 | 3.53 | 0.0005 *** |
| d.disc_rate.l1 | 0.4707601 | 0.0661962 | 7.11 | <0.0001 *** |
| d.cpi.l1 | -0.1080318 | 0.0384851 | -2.81 | 0.0054 ** |
| d.us_pop.l1 | 0.0000334 | 0.0003514 | 0.1 | 0.0924 |
| d.gdp.l2 | -0.0002258 | 0.0004353 | -0.52 | 0.6045 |
| d.disc_rate.l2 | -0.036785 | 0.0640857 | -0.57 | 0.5665 |
| d.cpi.l2 | -0.0239677 | 0.0388552 | -0.62 | 0.5379 |
| d.us_pop.l2 | -0.000229 | 0.0003523 | -0.65 | 0.516 |
| const | 0.2102579 | 0.1553768 | 1.35 | 0.1772 |
| sd1 | 0.0818089 | 0.0772542 | 1.06 | 0.291 |
| sd2 | 0.0092563 | 0.072679 | 0.13 | 0.8988 |
| sd3 | 0.0514729 | 0.0765898 | 0.67 | 0.5022 |
| exo1 | 0.0000195 | 0.0000116 | 1.68 | 0.0938 . |
| exo2 | 0.0000214 | 0.0000119 | 1.81 | 0.0718 . |
| exo3 | 0.0000199 | 0.000012 | 1.66 | 0.0978 . |
| $R^2$ | | | | 0.39 |
| F-statistic | | | | 8.32 on 14 and 251 DF |
| Sample Size | | | | 265 |

In Table 9, with short-run discount rate as the dependent variable, the first lags of GDP and discount rate are statistically significant at 0.1% significance level. First lag of CPI is significant @ 1% significance level. All the long-run co-integrating relationships are significant @ 10% significance level.
$R^2$ is 0.317.

Table 10 - d.cpi as the dependent variable

|  | Estimate | Std. Error | T-Statistic | Pr(>|t|) |
|---|---|---|---|---|
| d.gdp.l1 | 0.0010715 | 0.0007077 | 1.51 | 0.1312 |
| d.disc_rate.l1 | -0.1758159 | 0.1101163 | -1.6 | 0.1116 |
| d.cpi.l1 | 0.3007796 | 0.0640194 | 4.7 | <0.0001 *** |
| d.us_pop.l1 | 0.0004579 | 0.0005846 | 0.78 | 0.4342 |
| d.gdp.l2 | -0.0007335 | 0.0007241 | -1.01 | 0.312 |
| d.disc_rate.l2 | 0.0919693 | 0.1066054 | 0.86 | 0.3891 |
| d.cpi.l2 | -0.1301632 | 0.064635 | -2.01 | 0.0451* |
| d.us_pop.l2 | -0.0000372 | 0.000586 | -0.06 | 0.9494 |
| const | 0.7115997 | 0.2584667 | 2.75 | 0.0063 ** |
| sd1 | 0.032616 | 0.128511 | 0.25 | 0.7999 |
| sd2 | 0.0666619 | 0.1209004 | 0.55 | 0.5819 |
| sd3 | -0.1466519 | 0.1274059 | -1.15 | 0.2508 |
| exo1 | 0.0001414 | 0.0000193 | 7.32 | <0.0001 *** |
| exo2 | 0.0001455 | 0.0000197 | 7.38 | <0.0001 *** |
| exo3 | 0.000139 | 0.00002 | 6.97 | <0.0001 *** |
| $R^2$ | | | | 0.417 |
| F-statistic | | | | 12.8 on 14 and 251 DF |
| Sample Size | | | | 265 |

In Table 10, with short-run CPI as the dependent variable, the first and second lags of CPI are statistically significant @ 0.1% significance level and 5% significance level respectively. Intercept is significant @ 1% significance level. All the long-run co-integrating relationships are significant @ 0.1% significance level.
$R^2$ is 0.417.

Table 11 - d.us_pop as the dependent variable

| | Estimate | Std. Error | T-statistic | Pr(>\|t\|) |
|---|---:|---:|---:|---:|
| d.gdp.l1 | -0.01894 | 0.07222 | -0.26 | 0.79336 |
| d.disc_rate.l1 | 9.21673 | 11.23676 | 0.82 | 0.41286 |
| d.cpi.l1 | 4.07875 | 6.53282 | 0.62 | 0.53297 |
| d.us_pop.l1 | 0.51724 | 0.05965 | 8.67 | <0.0001 *** |
| d.gdp.l2 | -0.07145 | 0.07389 | -0.97 | 0.33445 |
| d.disc_rate.l2 | 5.52358 | 10.87849 | 0.51 | 0.61207 |
| d.cpi.l2 | 2.86255 | 6.59564 | 0.43 | 0.66466 |
| d.us_pop.l2 | 0.34287 | 0.0598 | 5.73 | <0.0001 *** |
| const | 88.91366 | 26.3751 | 3.37 | 0.00087 *** |
| sd1 | 65.04827 | 13.11384 | 4.96 | <0.0001 *** |
| sd2 | 186.98385 | 12.33721 | 15.16 | <0.0001 *** |
| sd3 | 132.88681 | 13.00106 | 10.22 | <0.0001 *** |
| exo1 | -0.00162 | 0.00197 | -0.82 | 0.41144 |
| exo2 | -0.00123 | 0.00201 | -0.61 | 0.54028 |
| exo3 | -0.00114 | 0.00204 | -0.56 | 0.57542 |
| $R^2$ | | | | 0.724 |
| F-statistic | | | | 47.1 on 14 and 251 DF |
| Sample Size | | | | 265 |

In Table 11, with short-run U.S population as the dependent variable, the first and second lags of population are statistically significant @ 0.1% significance level. The intercept, along with all the seasonal dummies are also statistically significant @ 0.1% significance level.
$R^2$ is 0.724.

*Model Performance*

For the four regression tables provided above with the summary of the outputs, we need to assess the performance of the model in order to examine how well the model is performing and whether we can improve the process or not. For measuring model performance, Root Mean Square Error Loss and Mean Absolute Percentage Error were calculated, which are based on the concept of measuring deviation of the predicted values of the output variable from the actual values. The following formulae were used –

$$R.M.S.E = \sqrt{\frac{\sum_{i=1}^{N}(y_i-\hat{y}_i)^2}{N}} \qquad M.A.P.E = \sum_{i=1}^{N}\frac{|y_i-\hat{y}_i|}{y_i}\times 100$$

Table 13 - Root Mean Square Error Loss for four linear equations

| Variable | Date N = 8 | Actual Value $y_i$ | Predicted Value $\hat{y}_i$ | R.M.S.E | M.A.P.E |
|---|---|---|---|---|---|
| GDP | Q2 2015 | 16461 | 16961 | 464 | 2.77% |
|  | Q3 2015 | 16528 | 17002 |  |  |
|  | Q4 2015 | 16548 | 17044 |  |  |
|  | Q1 2016 | 16572 | 17086 |  |  |
|  | Q2 2016 | 16664 | 17148 |  |  |
|  | Q3 2016 | 16778 | 17202 |  |  |
|  | Q4 2016 | 16851 | 17253 |  |  |
|  | Q1 2017 | 16903 | 17301 |  |  |
| Discount Rate | Q2 2015 | 0.75 | 1.41 | 0.44 | 40.51% |
|  | Q3 2015 | 0.75 | 1.39 |  |  |
|  | Q4 2015 | 0.83 | 1.31 |  |  |
|  | Q1 2016 | 1.00 | 1.15 |  |  |
|  | Q2 2016 | 1.00 | 1.04 |  |  |
|  | Q3 2016 | 1.00 | 0.94 |  |  |
|  | Q4 2016 | 1.08 | 0.85 |  |  |
|  | Q1 2017 | 1.33 | 0.71 |  |  |
| Consumer Price Index | Q2 2015 | 245 | 237 | 9.7 | 3.86% |
|  | Q3 2015 | 246 | 238 |  |  |
|  | Q4 2015 | 248 | 238 |  |  |
|  | Q1 2016 | 249 | 238 |  |  |
|  | Q2 2016 | 250 | 239 |  |  |
|  | Q3 2016 | 251 | 241 |  |  |
|  | Q4 2016 | 252 | 242 |  |  |
|  | Q1 2017 | 253 | 244 |  |  |
| U.S Population | Q2 2015 | 320972 | 325646 | 4649 | 1.44% |
|  | Q3 2015 | 321620 | 326276 |  |  |
|  | Q4 2015 | 322268 | 326906 |  |  |
|  | Q1 2016 | 322793 | 327434 |  |  |
|  | Q2 2016 | 323326 | 327973 |  |  |
|  | Q3 2016 | 323962 | 328605 |  |  |
|  | Q4 2016 | 324593 | 329235 |  |  |
|  | Q1 2017 | 325108 | 329762 |  |  |

Table 13 compares the actual values and the predicted values, and via forecast errors, computes R.M.S.E. The above out-of-sample forecasting was performed on time periods Q2 2015 to Q1 2017, providing a total of 8 quarters for forecasting purposes. R.M.S.E of model with short-run GDP as the dependent variable is 464. This suggests that on average, the model will over-predict or under-predict the true value of short-run GDP. On average, predicted short-run discount rate will be 0.44% higher or lower than the actual discount rate. Predicted CPI will be, on average, 9.7 less than or greater than the true value. The same interpretation can be used for short-run U.S population as well, which is in thousands.

The predicted values are obtained using the coefficients provided by the regression results. Although the models are not able to forecast with absolute precision, the rate at which the values vary as we go further ahead in time is in tandem with the rate at which actual values increase.

*Impulse Response Functions*

To trace out the responsiveness of the variables in the VECM built above and check how quickly each variable would return to its long-term equilibrium, impulse response functions (IRF) can be used. In this technique, a shock is introduced into one of the equations, in the system of equations, via the error term and then we can observe the change in impact of this shock on the other models which are part of the system. This helps in gauging the ramifications of this shock on the different variables in the model and also assists in analyzing the time it takes for the variables to return to steady-state.

Graph 10 Impulse Response – Short-Run GDP

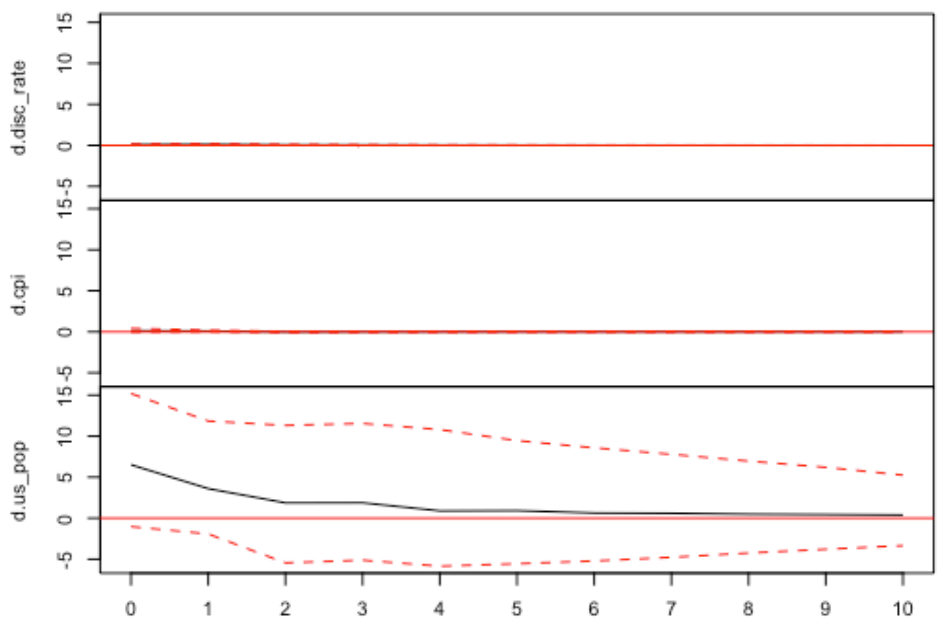

The above graph visually explains the dynamics of different variables when a shock is introduced into the model with short-run GDP as the dependent variable via the error term. This shock is inconsequential for the models with short-run discount rate and CPI as the dependent variables as the coefficients for GDP at different lag lengths are too small to have a significant impact. However, U.S population receives a positive shock at time period 0 and slowly returns to equilibrium. The speed with which the variables return to their equilibrium position is determined by the coefficient of the error correction component.

Graph 11- Impulse Response – Short-run Discount Rate

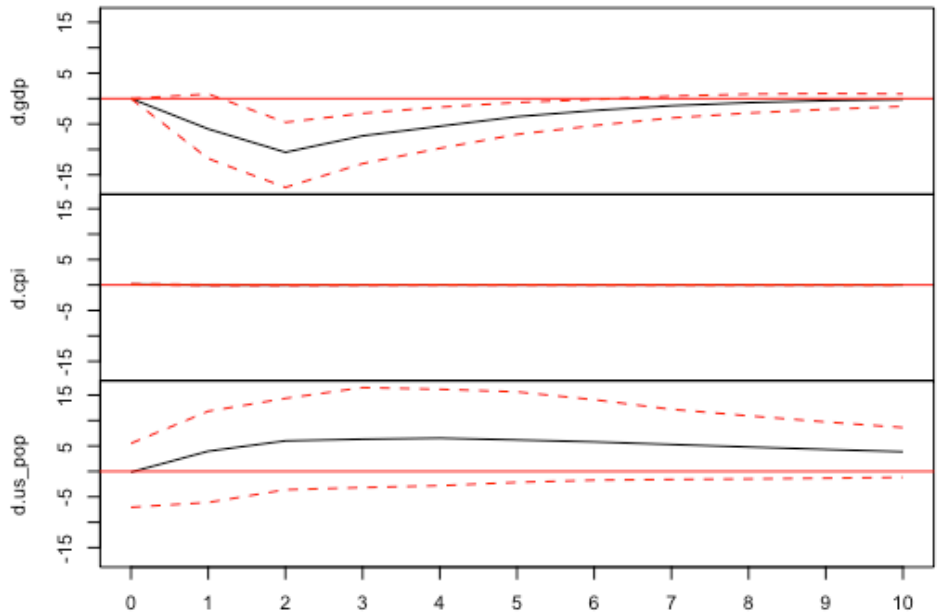

A shock introduced in the discount rate equation shows a negative correlation with the model which has short-run GDP as the response variable. On the other hand, this exogenous shock shows a positive relation with U.S population as population goes up slowly for some time, before returning back to its equilibrium level. CPI remains unchanged due to small coefficient representation of the discount rate. This impulse response function corroborates the hypothesis that a policy intended to have a particular impact might have some spillover effect, leading to changes that were not conceived earlier. Although decreasing short-run discount rate might increase GDP in the short-run, the above graph implies that it might also decrease U.S population, a policy effect that the government might want to cogitate over before proceeding.

When similar steps were followed for models with short-run CPI and U.S population as the dependent variables, none of the impulse response functions were significantly differently from zero and did not reveal any interesting relationship. Hence, they were not included in the final version of the paper.

*Forecast Error Variance Decomposition*

Variance decomposition tells us the proportion of change in one variable that can be attributed to other variables over time. It indicates the amount of information each variable contributes to the other variables in the autoregressive model and helps in identification of correlation between the variables at different time periods. In respective of forecasting, Variance decomposition of forecast errors determines how much of the forecast error variance of each of the variables involved can be explained by exogenous variables. The tables below provide the proportionate impact of different variables in determining the forecast error variance for each model for the 8 periods forecasted –

Table 14 - Variance Decomposition Matrix of Short-Run GDP

| d.gdp | d.disc_rate | d.cpi | d.us_pop |
|---|---|---|---|
| 1 | 0 | 0 | 0 |
| 0.99 | 0.0094 | 0.00014 | 0.000026 |
| 0.95 | 0.035 | 0.01489 | 0.000066 |
| 0.94 | 0.0465 | 0.01748 | 0.000065 |
| 0.93 | 0.0526 | 0.01771 | 0.000137 |
| 0.93 | 0.0552 | 0.01763 | 0.000211 |
| 0.93 | 0.0564 | 0.01759 | 0.000335 |
| 0.93 | 0.0568 | 0.01758 | 0.000459 |

The above table provides information about GDP, and informs about the fluctuations in short-run GDP that can be attributed to different variables. In time period 1, all the changes in GDP are due to its own change, and hence the proportionate change is equal to 1. This is the same as saying that GDP causes a 100% change in its own variance, and is therefore, independent of all the other variables. Furthermore, as we go ahead in time and look at the 8$^{th}$ period, GDP accounts for 93% fluctuation in its own variance, and short-run discount rate and short-run account for 5.6% and 1.7% of variance in GDP. Therefore, as we move further ahead in time, the variation in GDP becomes increasingly dependent on other involved variables.

Table 15 - Variance Decomposition Matrix of Short-Run Discount Rate

| d.gdp | d.disc_rate | d.cpi | d.us_pop |
|---|---|---|---|
| 0.055 | 0.94 | 0 | 0 |
| 0.112 | 0.86 | 0.023 | 0.000024 |
| 0.120 | 0.83 | 0.044 | 0.001352 |
| 0.125 | 0.82 | 0.053 | 0.002640 |
| 0.127 | 0.81 | 0.055 | 0.004219 |
| 0.128 | 0.81 | 0.055 | 0.005443 |
| 0.128 | 0.81 | 0.055 | 0.006441 |
| 0.128 | 0.81 | 0.055 | 0.007190 |

The above table can be interpreted in a similar manner as Table 14. However, an interesting characteristic of this table is that volatility in short-run discount rate is fairly dependent on short-run GDP and short-run CPI, with GDP and CPI accounting for 12.8% and 5.5% variation in the 8$^{th}$ time period.

Table 16 - Variance Decomposition Matrix of Short-Run Consumer Price Index

| d.gdp | d.disc_rate | d.cpi | d.us_pop |
|---|---|---|---|
| 0.033 | 0.038 | 0.93 | 0 |
| 0.045 | 0.037 | 0.92 | 0.0020 |
| 0.046 | 0.038 | 0.91 | 0.0029 |
| 0.046 | 0.038 | 0.91 | 0.0041 |
| 0.046 | 0.038 | 0.91 | 0.0049 |
| 0.046 | 0.038 | 0.91 | 0.0055 |
| 0.046 | 0.038 | 0.91 | 0.0060 |
| 0.046 | 0.038 | 0.91 | 0.0064 |

Fluctuations in short-run CPI in Table 16 seem to be decently dependent on short-run GDP and discount rate, with both the variables explaining 4.6% and 3.8% variation for the 8th forecasted periods.

Table 17 - Variance Decomposition Matrix of Short-Run U.S population

| d.gdp | d.disc_rate | d.cpi | d.us_pop |
| --- | --- | --- | --- |
| 0.0095 | 0.0000051 | 0.0006 | 0.99 |
| 0.0098 | 0.0027780 | 0.0025 | 0.98 |
| 0.0080 | 0.0070151 | 0.0044 | 0.98 |
| 0.0073 | 0.0107999 | 0.0050 | 0.98 |
| 0.0066 | 0.0141464 | 0.0052 | 0.97 |
| 0.0062 | 0.0167944 | 0.0053 | 0.97 |
| 0.0059 | 0.0188948 | 0.0054 | 0.97 |
| 0.0057 | 0.0205018 | 0.0054 | 0.97 |

Variation in forecasted errors for the model with short-run U.S population as the dependent variable seems to be independent of the explanatory variables involved in the model. All the other variables except U.S population together account for only 3% of the variation in U.S population.

## V. CONCLUSION

In this paper, an attempt was made to find the relationships between different macroeconomic variables which act as representatives of the economy and an effort was made to gauge their interdependence. Although such a study can be subjective based on the macroeconomic variables selected, the widely accepted economic indicators were used in assessing this relationship. Cointegrating relationships were identified between GDP, discount rate, CPI and U.S population, all of which are important economic indicators, using Engle-Granger cointegration test and Johansen cointegration test. The interdependence of the variables was analysed using a Vector Error Correction Model, with 3 steady-state relationships and short-run variables with a lag length of 2. The 4 models included in the system of equations forecasted values for 8 time periods and then the results were compared using R.M.S.E and M.A.P.E. Furthermore, the forecasts produced by 4 models which were part of the system of equations were extrapolated using IRFs and variance decomposition of forecast errors. This study provides quantifiable proof for significant interdependence of these economic indicators, proving that an economic policy can have more than just the desired effect and all the other relevant macroeconomic variables should also be taken into consideration when formulating an economic policy. Although this study is able to reveal some relationship between different models that were a part of the V.E.C.M, the variance decomposition matrix and impulse response function did not provide significant evidence to confirm the hypothesis. Limited size of the dataset and lack of correlation between the few economic indicators make the study subject to further analysis. Future studies that use more data along with better representatives of the economy can further support this investigation and provide greater insights onto the interdependence of various macroeconomic variables, giving policymakers more information about their interrelation so that more informed economic policies can be formulated in the future.